\documentclass[preprint,showpacs,preprintnumbers,amsmath,amssymb,showpacs,11pt]{revtex4}
\usepackage{graphicx}% Include figure files
\usepackage{dcolumn}% Align table columns on decimal point
\usepackage{bm}% bold math
\usepackage{amssymb}

\def\bea{\begin{eqnarray}}
\def\ena{\end{eqnarray}}

\begin{document}

\title{Cosmological Sakharov Oscillations and Quantum Mechanics of the Early
Universe}

\author{L. P. Grishchuk}
\affiliation{Sternberg Astronomical Institute, Lomonosov Moscow State 
University, Moscow 119899, Russia, and School of Physics and Astronomy, 
Cardiff University, Cardiff CF24 3AA, United Kingdom}

\date{\today}

%%%%%%%%%%%%%%%%%%%%%%%%%%%%%%%%%%%%%%%%%%%%%%%%%%%%%%%%%%%%%%%%%%%%%%%%%%

\begin{abstract}
This is a brief summary of a talk delivered at the Special Session of the
Physical Sciences Division of the Russian Academy of Sciences, Moscow, 25 May
2011. The meeting was devoted to the 90-th anniversary of the birth 
of A. D. Sakharov.
The focus of this contribution is on the standing-wave pattern of
quantum-mechanically generated metric (gravitational field) perturbations
as the origin of subsequent Sakharov oscillations in the matter power
spectrum. Other related phenomena, particularly in the area of gravitational
waves, and their observational significance are also discussed.
\end{abstract}

%%%%%%%%%%%%%%%%%%%%%%%%%%%%%%%%%%%%%%%%%%%%%%%%%%%%%%%%%%%%%%%%%%%%%%%%%%


\maketitle

%%%%%%%%%%%%%%%%%%%%%%%%%%%%%%%%%%%%%%%%%%%%%%%%%%%%%%%%%%%%%%%%%%%%%%%%%%
%%%%%%%%%%%%%%%%%%%%%%%%% MAIN TEXT %%%%%%%%%%%%%%%%%%%%%%%%%%%%%%%%%%%%%%
%%%%%%%%%%%%%%%%%%%%%%%%%%%%%%%%%%%%%%%%%%%%%%%%%%%%%%%%%%%%%%%%%%%%%%%%%%

%%%%%%%%%%%%%%%%%%%%%%%%%%%%%%%%%%%%%%%%%%%%%%%%%%%%%%%%%%%%%%%%%%%%%%%%%%%%%%%%%%%%%%%%%%%

\section{\label{sec:history}Sakharov's first cosmological paper}

The ideas and results of Andrei Sakharov's remarkable paper \cite{Sakh65} 
have influenced the course of cosmological research and are still
in the centre of theoretical and observational studies. The title of his paper
was ``The initial stage of an expanding universe and the appearance of a
nonuniform distribution of matter". The paper was submitted to ZhETF on
2 March 1965, that is, in the days when not only the existence of the cosmic
microwave background radiation (CMB) was not yet established, but even the
nonstationarity of the Universe was still debated. Right the second sentence
of the Abstract says: ``It is assumed that the initial inhomogeneities arise
as a result of quantum fluctuations of cold baryon-lepton matter at densities
of the order of $10^{98}$ baryons/cm$^3$. It is suggested that at such
densities gravitational effects are of decisive importance in the equation of
state...".

In what follows, we discuss recent attempts to explain the appearance 
of cosmological perturbations (density
inhomogeneities, gravitational waves, and possibly rotational perturbations)
as a result of quantum processes. In our approach, the perturbations
arise as a consequence of superadiabatic (parametric) amplification of
quantum-mechanical fluctuations of the appropriate degrees of freedom of the
gravitational field itself. So, for us, gravity is of decisive importance
not so much because of its contribution to the equation of state of the
primeval matter, but because the gravitational field (metric) perturbations
are the primary object of quantization. Nevertheless, it must be stressed
that the mind-boggling idea suggesting that something microscopic and
quantum-mechanical can be responsible for the emergence of fields and observed
structures at astronomical scales was first formulated and partially
explored in Sakharov's paper.

A considerable part of the paper \cite{Sakh65} is devoted to the evolution
of small density perturbations, rather than to their origin.
The spatial Fourier component of the relative density perturbation
is denoted $z_{\kappa}(t)$, where $\kappa$ is
a wavenumber. The function $z_{\kappa}(t)$ satisfies a second-order
differential equation, numbered in the paper as Eq.(15), which follows from
the perturbed Einstein equations. The calculation leading to the phenomenon
which was later named the Sakharov oscillations is introduced by the following
words:
\newline
{\it Yu. M. Shustov and V. A. Tarasov have at our request solved Eq.(15), with
the aid of an electronic computer, for different values of $\kappa$. The
calculations were made for the simplest equation of state, satisfying
$\epsilon = nM$ with $n^{1/3} << M$ and $\epsilon = A n^{4/3}$ with
$n^{1/3} >> M$ (A is a constant $\sim 1$)
\[
\epsilon = n\left(M^2 + A^2 n^{2/3}\right)^{1/2}. ~~~~~~~~~(16)
\]}

In the paper \cite{Sakh65}, the quantity $n$, $n= 1/ a^{3}(t)$, is the
particle number density, $\epsilon$ is the energy density in the rest
frame of the material, and $p=n d\epsilon/dn - \epsilon$ is the pressure.
Obviously, the interpolating formula (16) describes the transition from
the relativistic equation of state $p= \epsilon/3$, applicable at
early times of evolution and relatively large $n$, to the nonrelativistic
equation of state $p=0$ valid at small $n$ and late times. During the
transition, the speed of sound decreases from $c_s = c/{\sqrt 3}$ to
$c_s = 0$.

It is important to realize that the physical nature of the discussed
transition from $p=\epsilon/3$ to $p=0$ can be quite general. Being guided by
physical assumptions of his time, Sakharov speaks about cold baryon-lepton
matter, degenerate Fermi gas of relativistic noninteracting particles, and so
on. But it is important to remember that the perturbed Einstein equations,
such as Eq.(15), do not require knowledge of microscopic causes of elasticity
and associated speed of sound. Gravitational equations operate with the
energy-momentum tensor of the material and its bulk mechanical properties,
such as the energy density, pressure, and the link between them, the 
equation of
state. These are postulated by Eq.(16), and one can now think of the
results of the performed calculation as a qualitative model of what can
happen in other transitions. For example, in a transition
from the fluid dominated by photon gas with the equation of
state $p= \epsilon/3$ to the fluid dominated by cold dark matter (CDM) with
the equation of state $p=0$.

For simple models of matter, such as $\epsilon = n^{\gamma}$ and
$p=(\gamma-1) \epsilon$, Eq.(15) can be solved in elementary functions.
Sakharov writes:
\newline
{\it When $\gamma=const$, the solution of
this equation is expressed in terms of Bessel functions; for example, when
$\gamma= 4/3$ we have increasing and decreasing solutions of the form
($\theta \sim t^{1/2} \kappa$)
\[
z\propto \left\{ \begin{array}{cl}
\cos \theta - \theta^{-1} \sin \theta, \\
\sin \theta + \theta^{-1} \cos \theta.
                \end{array}
\right. \]}

Indeed, these are the well known solutions for $z_{\kappa}(t)$ in the
$p=\epsilon/3$ medium. The general solution to Eq.(15) is a
linear combination of these two branches with arbitrary (in general,
complex) coefficients. The first
solution can be called increasing and the second decreasing, because at very
small $\theta$ they behave as $\theta^2$ and $\theta^{-1}$, respectively.
At later times, not long before the transition to the $p=0$
regime, the functions $z_{\kappa}(t)$ represent ordinary acoustic waves with
the oscillatory time dependence $\cos \theta$ and $\sin \theta$. One does not
learn anything new from matching the increasing/decreasing part of a solution
to the oscillatory part of the same solution; the general solution is already
given by the formula above. At the $p=0$ stage the solutions for
$z_{\kappa}(t)$ do not oscillate as functions of time; they are power-law
functions of $t$.

The crucial observation in the Sakharov paper is contained in the following
quotation:
\newline
{\it The function $a(t)$ can be obtained in the case of Eq.(16) analytically
(Shustov). Shustov and Tarasov find, by integrating (15) the limiting value
as $t \rightarrow \infty$ of the auxilliary variable
\[
\zeta = z\left(1+ a^2 M^2/A^2\right)^{-1/2},
\]
putting $d\zeta/dt=dz/dt \sim z_0$ as $t \rightarrow 0$. It is obvious that
$\zeta(\infty) \propto z_{0} B$. \\
~~~~~In accordance with the results of the sections
that follow, we put $z_0 \sim \kappa$. $\zeta(\infty)$ is a function of the
parameter $A^{1/2} \kappa$. This function is oscillating and sign-alternating,
but attenuates rapidly with increasing $\kappa$.}

The last sentence of this quotation is a surprising statement of incredible
importance. It says that well after the transition to the $p=0$
regime ($t \rightarrow \infty$) the density fluctuation $z_{\kappa}(t)$
becomes an oscillating and sign-alternating function of the wavenumber
$\kappa$. The square of this function is what can be called a power spectrum.
Sakharov uses $z_{\kappa}(t)= z_{0 \kappa}/{\dot a}^2$  at the very early
times and  takes $z_{0 \kappa}$ as $z_0 \sim \kappa$ from his
quantum-mechanical considerations. So, it is stated that the initial smooth
power spectrum $z_{0\kappa}^2$  transforms into an oscillatory final
power spectrum which has a series of zeros and maxima at some specific
wavenumbers $\kappa$. If one imagines that in the era before the transition
to the $p=0$ regime the field of
sound waves was represented by a set of harmonic oscillators with different
frequencies, then the claim is that well after the transition some oscillators
will find themselves ``lucky", in the sense that they appear in the maxima
of the resulting power spectrum, while others - ``unlucky", because they are at
the zeros of the resulting power spectrum.

Certainly, such a striking conclusion cannot be
unconditionally true. After all, a computer can be asked to
make a similar calculation, but backwards in time. In this calculation, one can
postulate a smooth power spectrum at the late $p=0$ stage and evolve the
spectrum back in time to derive the functions $z_{\kappa}(t)$ at the early
$p=\epsilon/3$ stage. The derived functions
will not coincide with what was taken as initial conditions in the original
calculation \cite{Sakh65}, but such new initial conditions are possible in
principle. By construction, these new initial conditions would not lead to
the final power spectrum oscillations. On the other hand, if the
oscillations do arise from physically justified initial conditions,
then this is an extremely important phenomenon. It dictates the appearance
of a periodic structure in Fourier space (a ``standard ruler" with
characteristic spatial scales) which
can be recognized in observations and can be used as a tool for other
measurements.

The point of this remark
is to stress that, as will be argued below in more detail, the initial
conditions leading to Sakharov oscillations are inevitable, if the primordial
cosmological perturbations were indeed generated quantum-mechanically.

The oscillatory transfer function $B(\kappa)$ participates in further
calculations \cite{Sakh65}, but it takes quite a modest role there.
Sakharov himself did not elaborate on the discovered phenomenon in
later publications. However, it seems to me that he was perfectly well aware
of the importance of his observation, and he attentively followed subsequent
developments. Some evidence for this will be given in sec.\ref{sec:quantum}.

It was Ya.B.Zeldovich who assigned significant value to the discovered
oscillations and named them the Sakharov oscillations. In conversations,
at seminars, in papers with R.A.Sunyaev, A.G.Doroshkevich, and in a
book with I.D.Novikov, Zeldovich discussed the physics of the
phenomenon and its possible observational applications. Zeldovich and
coauthors deserve
credit for seeing the relevance of Sakharov's work for their own studies
and for mentioning his paper. For example, one of the first papers
on the subject in the context of a ``hot" model of the Universe \cite{SZ70}
remarks: ``at a later stage of expansion the amplitude of
density perturbations turns out to be a periodic function of a wavelength
(mass). Such a picture was previously obtained by Sakharov (1965)
for a cold model of the Universe". And more \cite{SZ70}: ``The picture
presented above is only a rough approximation since the phase
relations between density and
velocity perturbations in standing waves in an ionized plasma were not
considered. As mentioned in the introduction, Sakharov (1965) showed that the
amplitude of perturbations of matter at a later stage when pressure does not
play a role (in our case after recombination) turns out to be a periodic
function of wavelength". Zeldovich and Novikov \cite{ZN83} discuss the
phenomenon at some length and note that
``The distribution of astronomical objects with respect to mass will thus
reflect the Sakharov oscillations in a very smoothed-out form only. It is
possible that they may not be noticed in a study of the mass spectrum".
Fortunately, as we shall see below, there was a significant observational
progress in revealing Sakharov oscillations.

The parallel to \cite{SZ70} and more detailed paper by P.J.Peebles
and J.Yu \cite{PY70} explicitly presents in Fig.5 a modulated spectrum, with
maxima and zeros, and mentions the relevance of the ``first big peak
in Fig.5" to the future experimental searches of irregularities in the
microwave background radiation. The spectral modulation was derived as a result
of numerical calculations. The later private correspondence on the physical
interpretation of oscillations inevitably ended up with ``lucky" and
``unlucky" oscillators \cite{Peebles90}: ``The Sakharov oscillations you
mention also were considered by Jer Yu and me (a few years after Sakarov)....
Here there truly are modes that are unlucky, in the sense that they carry
negligible energy".

To better understand the Sakharov oscillations, as well as other closely
related phenomena, we have to make some formalization of the problem. We will
do this in the next section. Before that, it is interesting to note, as a
side remark, that in course
of his quantum-mechanical considerations Sakharov discusses the ``initial
stage of the expansion of the universe", and in particular with the scale
factor $a = e^{\lambda t}$ as $t \rightarrow -\infty$. He found this evolution
in two cases, $c$ and $d$, out of the four considered. This type of the scale
factor is now advertized as inflation. However, Sakharov himself was
sceptical about cases $c$ and $d$. He finds arguments against them and
concludes: ``For these reasons we turn to curves $a$ and $b$". (Criticism
of contemporary inflationary claims can be found in \cite{GrWhel},
\cite{GrComm}.)

\section{\label{sec:fields}Wave-fields of different nature in time-dependent
environments}

The main physical reason behind Sakharov oscillations, and indeed behind
many other similar phenomena, is the time-dependence of the parameters
characterizing the environment in which a wave-field is given. This
can be a changing speed of sound, or a changing background gravitational
field, or all such factors together.
In cosmology, the central object is the gravitational field (metric)
perturbations. Other quantities, such as fluctuations in density and velocity
of matter (if they are present; we recall that they are absent in the case of
gravitational waves), are calculable from the metric perturbations via the
perturbed Einstein equations. It is only in special conditions and for
relatively short-scale variations that the gravitational field perturbations
can be neglected.

The gravitational field perturbation $h_{ij}$ is defined by
\begin{eqnarray}
ds^{2} = -c^{2} dt^{2} + a^{2}(t)(\delta_{ij} + h_{ij}) dx^idx^j=
a^{2}(\eta)\left[ - d\eta^{2} + (\delta_{ij} + h_{ij}) dx^idx^j\right].
\label{FRWmetric}
\end{eqnarray}
For each of the three types of cosmological perturbations (density
perturbations, gravitational waves, and rotational perturbations) the
field $h_{ij}$ can be expanded over spatial Fourier modes with wavevectors
${\bf n}$:
\begin{eqnarray}
\label{hij}
h_{ij} (\eta ,{\bf x})
= \frac{\cal C}{(2\pi )^{3/2}} \int_{-\infty}^\infty d^3{\bf n}
  \sum_{s=1, 2}~{\stackrel{s}{p}}_{ij} ({\bf n})
   \frac{1}{\sqrt{2n}}
\left[ {\stackrel{s}{h}}_n (\eta ) e^{i{\bf n}\cdot {\bf x}}~
                 {\stackrel{s}{c}}_{\bf n}
                +{\stackrel{s}{h}}_n^{\ast}(\eta) e^{-i{\bf n}\cdot {\bf x}}~
                 {\stackrel{s}{c}}_{\bf n}^{\dag}  \right].
\end{eqnarray}

The power spectrum (variance) of a given field is a quadratic combination of
the field averaged over space, or over known classical probability
density function, or over known quantum-mechanical state. In all cases,
one arrives at the expression of the following structure
\begin{eqnarray}
\left<0\right|h_{ij}(\eta,{\bf x})h^{ij}(\eta,{\bf
x})\left|0\right> =
\frac{\mathcal{C}^{2}}{2\pi^{2}}\int\limits_{0}^{\infty}
~n^{2}\sum_{s=1,2}|\stackrel{s}{h}_n(\eta)|^{2}\frac{dn}{n}.
\label{meansq}
\end{eqnarray}
The quantity
\begin{eqnarray}
h^{2}(n,\eta) = \frac{\mathcal{
C}^{2}}{2\pi^{2}}n^{2}\sum_{s=1,2}|\stackrel{s}{h}_n(\eta)|^{2}
\label{gwpower}
\end{eqnarray}
is called the metric power spectrum. At each instance of time, the metric power
spectrum is determined by the absolute value of the (in general, -- complex)
gravitational mode functions $\stackrel{s}{h}_n(\eta)$. (We often suppress
the index $s,~s=1,2$, which marks two polarisiation states present in metric
perturbations of each type
of cosmological perturbations.) For calculation of power spectra of other
quantities participating in the problem, one has to expand these quantities
as in Eq.(\ref{hij}) and then use their mode functions in expressions for their
power spectra, similar to Eq.(\ref{gwpower}).

The gravitational mode functions $\stackrel{s}{h}_n(\eta)$, as well as mode
functions of other quantities participating in our problem, satisfy one or
another version of the second-order differential ``master equation" \cite{BG}
\begin{equation}
\label{master}
f^{\prime\prime} + f \left[n^2 \frac{c_s^2}{c^2} - W(\eta) \right] = 0 \ \,
\end{equation}
where the ``speed of sound" $c_s$ and the ``potential" $W(\eta)$ are, in
general, functions of time. In particular, the Sakharov mode functions
$z_{\kappa}(t)$ for density perturbations obey a specific equation of this
kind (written in the $t$-time). And the above-quoted Sakharov solution, for
$\gamma=4/3$, expressed in terms of Bessel functions with the
argument $\theta$ is a particular case in which $c_s = c/{\sqrt 3}$,
whereas $W(\eta)$ is a simple function of the scale factor $a(\eta)$.
Gravitational wave equations are also equations of this form with
$c_s = c$.

Two linearly-independent high-frequency solutions (i.e. solutions 
of ``master equation" (\ref{master}) without $W(\eta)$ and with
$c_s=const$) are usually taken as $f_{n}(\eta) = e^{{\pm} in(c_s/c)\eta}$. 
If these mode
functions $f_{n}(\eta)$ represent sound waves not long before the transition
to the $p=0$ regime, then using them for calculation of
the power spectrum one would find $|f_{n}|^2=1$ and, hence, the absence of
oscillations in the power spectrum of density perturbations. Therefore, we do 
not expect any segregation into ``lucky" and ``unlucky" oscillators in
the post-transition era. The general decomposition (\ref{hij}) should be
looked at more closely.

The general high-frequency solution to Eq.(\ref{master}) (for simplicity, we
set temporarily $c_s/c = 1$) is $f_n(\eta) = A_n e^{-in\eta} +
B_n e^{in\eta}$, where complex coefficients $A_n,~B_n$ are in general
arbitrary functions of $n$. The ${\bf n}$-mode of the field
\[
h_{\bf n} (\eta ,{\bf x}) =f_n (\eta ) e^{i{\bf n}\cdot {\bf x}} +
f_n^{*}(\eta) e^{-i{\bf n}\cdot {\bf x}}
\]
is a sum of two waves traveling in opposite directions with arbitrary
amplitudes and arbitrary phases. One particular traveling wave is chosen by
setting $|A_n|=0$ or $|B_n|=0$. In contrast, the choice $|A_n|=|B_n|$ makes
the field a standing wave, that is, a product of a function of $\eta$ and a
function of ${\bf n}\cdot {\bf x}$:
\[
h_{\bf n} (\eta ,{\bf x}) =4 \rho_A \cos\left(n\eta +
\frac{\phi_B-\phi_A}{2}\right) \cos \left({\bf n}\cdot {\bf x} +
\frac{\phi_B+\phi_A}{2}\right),
\]
where we have used $A_n=\rho_{A_n} e^{i\phi_{A_n}}$,
$B_n=\rho_{B_n} e^{i\phi_{B_n}}$ without the label $n$.

The power spectrum of the general solution is
\[
|f_n|^2 = \rho_A^2 + \rho_B^2 +2\rho_A\rho_B \cos(2n\eta+\phi_B-\phi_A).
\]
Clearly, for a given moment of time $\eta$, the spectrum is a modulated
function of $n$. For the modulation to take form of a strict
periodic oscillation, the phase $\phi_B-\phi_A$ should be a linear
function of $n$. The oscillations vanish for traveling waves and have
the maximal depth, up to the appearance of zeros, for standing waves. In
principle, $\rho_A$ and $\rho_B$ could themselves be complicated functions
of $n$, but for the moment we do not consider this possibility.
\begin{figure*}[t]
\begin{center}
\includegraphics[width = 14cm]{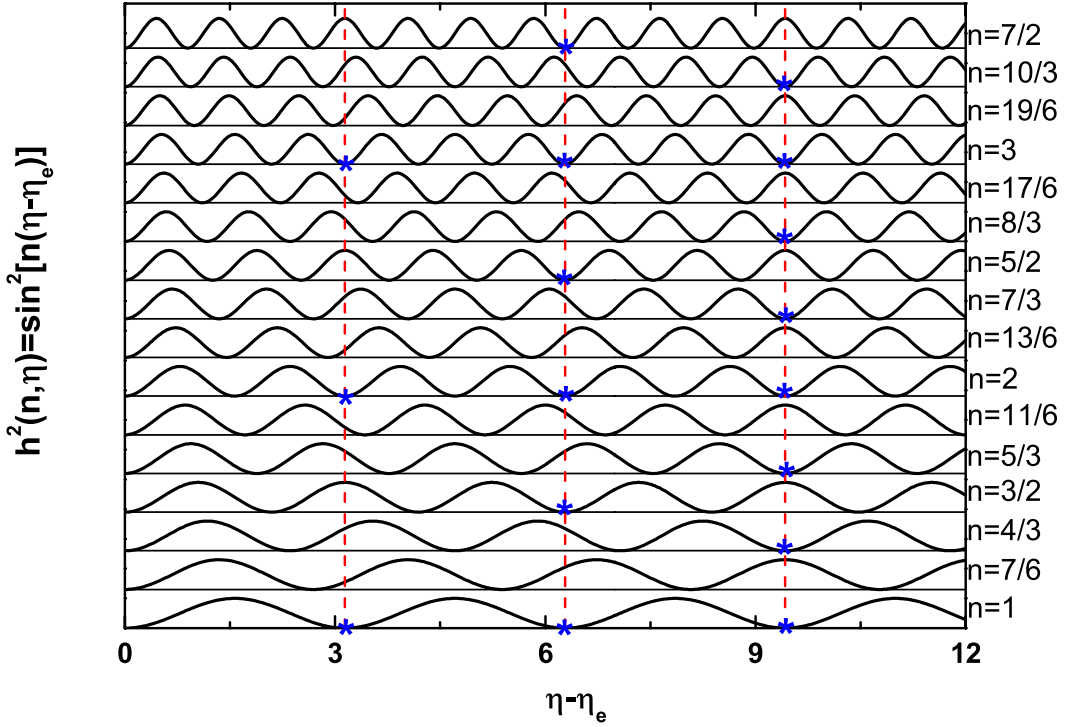}
\end{center}
\caption{A model spectrum of the pre-transition wave-field with moving
(proliferating) zeros.}\label{Sakh5.pdf}
\end{figure*}

For illustration, we show in Fig.\ref{Sakh5.pdf} a model spectrum
$h^2(n, \eta) = \sin^2[n(\eta - \eta_e)]$ ($\eta_e=const$) plotted for a
discrete set of wavenumbers $n$. The zeros in the spectrum, marked by blue
stars, move and prolifirate in the course of time, in the sense that they
gradually arise at new frequencies, and the distance between them decreases.
The moving zeros and moving maxima will be inherited and fixed (possibly,
with a phase shift) in the power spectrum at the  $p=0$ stage after 
the transition.

Indeed, the general solution of Eq.(\ref{master}) after the transition is
$f_n(\eta) = C_n +D_n \eta$. It is the coefficients $C_n, D_n$ that
become oscillatory functions of $n$. The moving features
become fixed features at some particular wavenumbers, thus defining the
``lucky" and ``unlucky" oscillators. If the transition can be approximated
as a sharp event occuring at some $\eta_{eq}$, then by joining the general
solutions for the function $f_n(\eta)$ and its first time 
derivative $f^{\prime}_n(\eta)$
at $\eta = \eta_{eq}$, we find for the coefficient in the growing solution
\[
|D_n|^2 =n^2(c_s/c)^2 [\rho_A^2 + \rho_B^2 - 2\rho_A\rho_B
\cos(2n(c_s/c)\eta_{eq}+\phi_B-\phi_A)].
\]
Obviously, there are no final spectrum modulations if the incoming field
consists of traveling waves ($\rho_A =0$ or $\rho_B =0$), and the 
modulations
have maximal depth if the waves are standing ($\rho_A =\rho_B$). 
The relevant
set of maxima is determined by the set of $n$ where the function
$\sin^2[(c_s/c)n\eta_{eq}+(\phi_B-\phi_A)/2 ]$ has a maximum, starting
from $(c_s/c)n\eta_{eq}+(\phi_B-\phi_A)/2= \pi/2$. The smallest
$n$ and hence the largest spatial scale $\lambda = 2\pi a(\eta)/n$ is expected
to be the most pronounced observationally. For such long wavelengths, the
metric perturbations cannot be generically neglected. Note that
if the $p=0$ post-transition medium is CDM, then there must be oscillations 
in the CDM power spectrum.

One can see that it is only a very high degree of organization of the field
before the transition, -- standing waves with phases proportional to $n$, --
that can lead to the emergence of periodic Sakharov oscillations in the
post-transition pressureless matter and in the associated metric
perturbations.

The power spectra of cosmological fields in the recombination era
determine the angular
power spectrum of cosmic microwave background anisotropies observed
today. Depending on whether the perturbations are realised as traveling or
standing waves, the CMB spectra will be strongly different. This is
best illustrated with the help of
gravitational waves. In the case of gravitational waves only gravity is
involved, so one should not worry about the ``acoustic physics" and the role
of various matter components. The decoupling of photons from baryons at
the last scatering surface $\eta = \eta_{dec}$ has no effect on
gravitational waves themselves, but for the photons it is very important in
which gravitational field they start their journey and propagate.

\begin{figure*}[t]
\begin{center}
\includegraphics[width = 12cm, height=16cm]{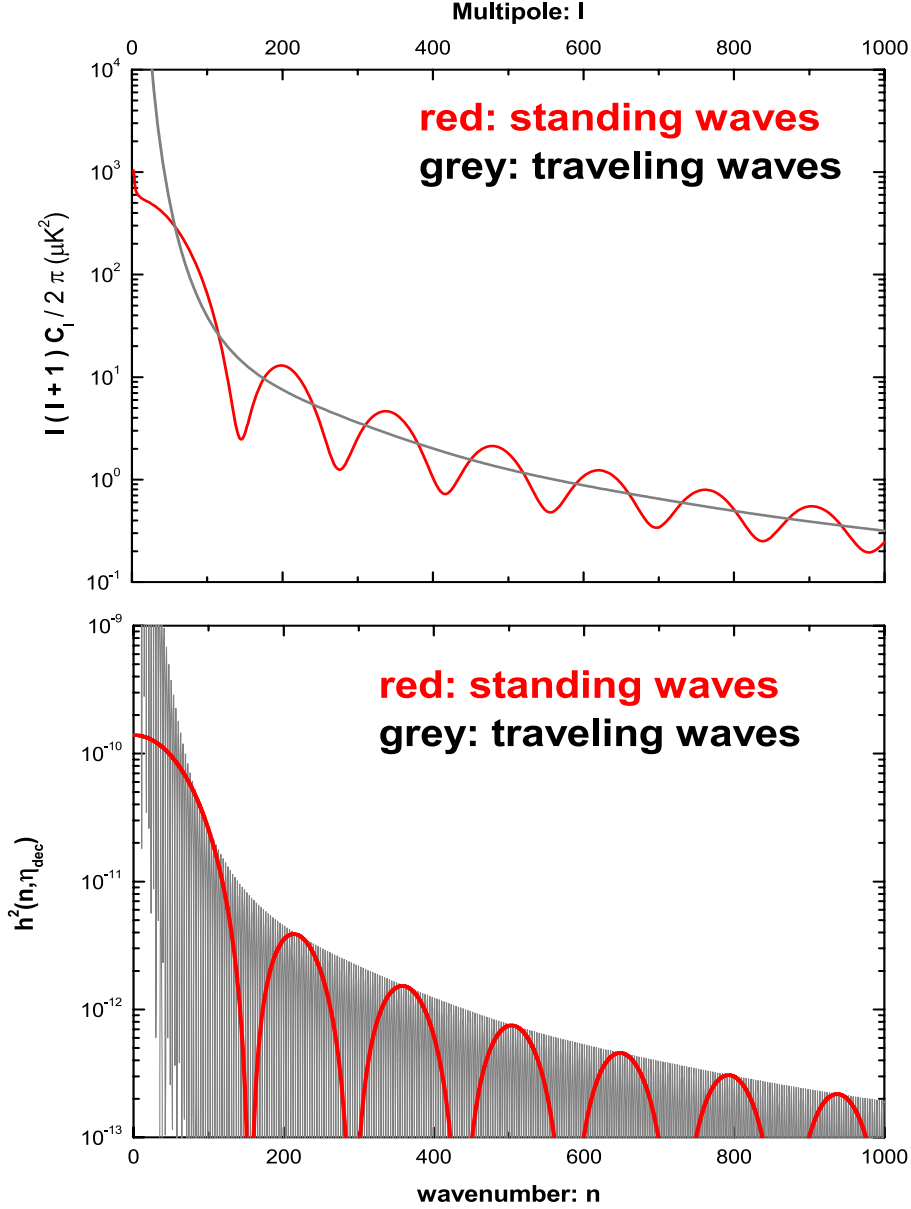}
\end{center}
\caption{Angular power spectra of CMB temperature anisotropies (upper panel)
generated by power spectra of standing or traveling gravitational
waves (lower panel)}

\label{Sakh4.pdf}

\end{figure*}

In Fig.\ref{Sakh4.pdf} we show two power spectra of gravitational waves
given at $\eta = \eta_{dec}$ and two corresponding CMB temperature spectra
caused by them (more details in \cite{BG}). The red (wavy) line describes the
physical spectrum formed by (quantum-mechanically generated) standing waves,
whereas the grey (smooth) line shows the alternative background formed by
traveling waves.
The power spectrum of the alternative background was chosen to be an envelope
of the physical one, so that the broad-band powers in the two spectra are
approximately equal, except at very small $n$'s. The CMB spectra are placed
right above the underlying gravitational wave spectra in order to demonstrate
the almost one-to-one correspondence between their features in $n$-space and
$\ell$-space. A similar correspondence holds for the power spectrum of first
time-derivative of the $h_{ij}$ field and CMB polarization spectra for which
it is responsible \cite{BGP}. It is important to note that the planned new
sensitive measurements of CMB polarization and temperature (e.g. \cite{Spider})
may be capable of identifying the first cycle of oscillations in the physical
gravitational-wave background.

\section{\label{sec:observ}Current observations of oscillations in power
spectra of matter and CMB.}

It should be clear from the discussion above that the Sakharov oscillations
are not trivial acoustic waves in relativistic plasma. Such acoustic waves, 
expressing the variability of physical quantities in space and time, 
always exist, in the sense that
they are general solution to the density fluctuation equation. The Sakharov
oscillations are something much more subtle. They are the variability in the
post-transition power spectrum, that is, oscillations in Fourier space. At late
times, the oscillatory shape of the matter power spectrum remains fixed. The
oscillations define the preferred wavenumbers and spatial scales, in agreement
with the standing-wave pattern of the pre-transition field.

Oscillations in the final power spectrum do not arise simply as a result of a
``snapshot" of oscillations in the baryon-photon fluid or as an ``impression"
of acoustic waves in the hot plasma of the early universe onto the matter
distribution. And they are neither the result of the propagation of spherical
sound waves up to the ``sound horizon" before recombination, nor the result
of ``freezing out" of traveling sound waves at decoupling. The event
when the plasma becomes transparent can make the Sakharov oscillations
visible, but this is not the reason why they exist. Periodic
structures in the final power spectrum arise only if the sound waves in
relativistic plasma (as well as the associated metric perturbations) are
standing waves with special phases. The oscillations in the power spectrum
do not arise at all if the sound waves are propagating. It is also clear
from the discussion above that the phenomenon of oscillations is not specific
to baryons. The oscillations are present, for example, in the power spectrum
of metric perturbations accompanying matter fluctuations and in gravitational
waves.

It appears that actual observations have revealed convincing traces of
Sakharov oscillations in the distribution of galaxies. Existing and
planned surveys concentrate on the distribution of luminous matter (baryons)
and therefore the spectral features are often called  
the baryon acoustic oscillations (BAO). The structures in the power 
spectrum are Fourier-related to the spikes in the two-point spatial 
correlation function. Both characteristics have been measured
in galaxy surveys, (e.g. \cite{Cole05}, \cite{Eis05}, \cite{Perc07},
\cite{corrf3}; the last citation contains many references to previous work.)

Of course, the ideal picture of standing waves in the early plasma
is blurred by the multi-component nature of cosmic fluid and by the variety
of astrophysical processes happenning on the way to the observed spatial
distribution of nonrelativistic matter. This makes the oscillatory
features much smoother and much more difficult to identify. Moreover, the
measurement of our own particular realization of the inherently random field
is only an estimate of the theoretical, statistically averaged, power
spectrum, such as Eq.(\ref{gwpower}). Nevertheless, the impressive
observations of recent years gave significant evidence of the existence of
Sakharov oscillations.

A similar situation takes place in the study of CMB temperature and
polarization. The difference between smooth and oscillatory underlying
spectra for the ensuing CMB anisotropies was illustrated by gravitational
waves in Fig.\ref{Sakh4.pdf}. Density perturbations are more complicated 
because they include the individual power spectra of fluctuations 
in matter components, the velocity of the fluid which emits and 
scatters CMB photons (the velocity and the
associated Doppler terms require careful definitions), and gravitational 
field perturbations. Surely, the observed peaks and dips 
in CMB temperature angular
spectrum $C_{\ell}^{TT}$, now measured up to high multipoles
$\ell$ \cite{ACT}, are a reflection of oscillations in the underlying power
spectra at the time of decoupling $\eta_{dec}$. (A link with the phenomenon
of Sakharov oscillations, in some generalized sense, was mentioned
in \cite{Jorg95}, \cite{BG}.) It is very likely that the oscillations in
$C_{\ell}^{TT}$ at relatively high $\ell$'s are a direct reflection of
standing-wave pattern of density variations in baryon-electron-photon plasma
itself, so they are ``acoustic" signatures. In contrast, the structures at
the lowest $\ell$'s are probably having a considerable contribution from
the pre-transition metric perturbations, which were inherited at the time of
transition $\eta_{eq}$, mostly by the gravitationally dominant cold dark
matter, so these structures are more like ``gravitational" peaks and
dips \cite{BG}. [The current cosmological literature emphasizing the
``acoustic" side of the problem incorrectly claims that there should not be 
oscillations in the power spectrum of CDM.]

\begin{figure*}[t]
\begin{center}
\includegraphics[width = 10cm,angle=-90]{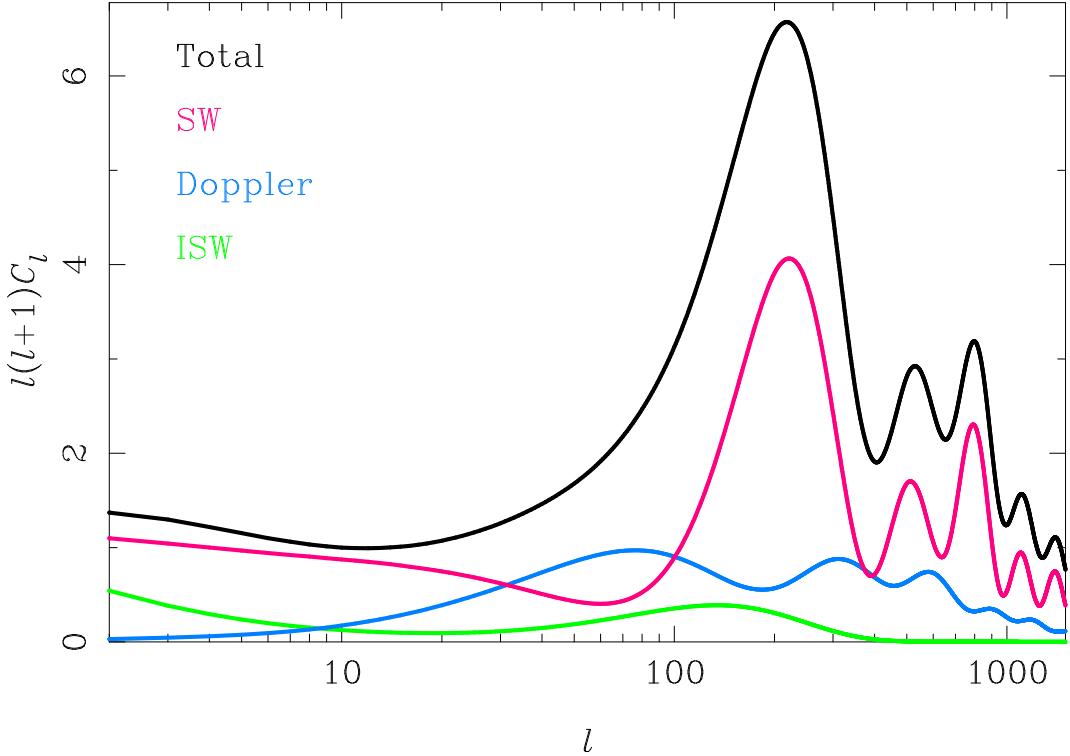}
\end{center}
\caption{Various contributions to CMB temperature anisotropies
(from \cite{Chal04}).}\label{Chal.pdf}
\end{figure*}

It should be remembered,
however, that the decomposition of the total CMB signal into different
contributions is not unambiguous, and the interpretation may depend on
coordinate system (gauge) chosen for the description of fluctuations.
In the so-called Newtonian gauge the decomposition of the total signal is
presented in Fig.\ref{Chal.pdf}, taken from \cite{Chal04}. The dominating
SW contribution (SW stands for Sachs-Wolfe) is a combination of variations of
the metric and photon density.

We can make the following intermediate conclusions. First, for the Sakharov
oscillations to appear in the final matter power spectrum, they must be
encoded from the very beginning in the power spectrum of primordial
cosmological perturbations, as a consequence of standing waves. Therefore, 
the Sakharov oscillations must have truly primordial origin 
(quantum-mechanical, as we argue below). Second, the very existence of 
periodic structures in the power spectra of matter and CMB is not a lesser 
revelation about the Universe than those future discoveries that will
hopefully be made with the help of
these ``standard rulers". In particular, in the case of data from galaxy
surveys, it is important to be sure that we are dealing with manifestations
of Sakharov oscillations, and not with something else. If they are Sakharov
oscillations, then the phases were remembered for 13 billion years. Third,
at some elementary level the Sakharov oscillations can be tested in laboratory
conditions. This is a difference in fates of traveling and standing waves in
a medium in which the sound speed changes from large values to zero. It would
be useful to perform this experimental demonstration.

\section{\label{sec:quantum}Quantum mechanics of the very early Universe.}

\begin{figure*}[t]
\begin{center}
\includegraphics[width = 12cm]{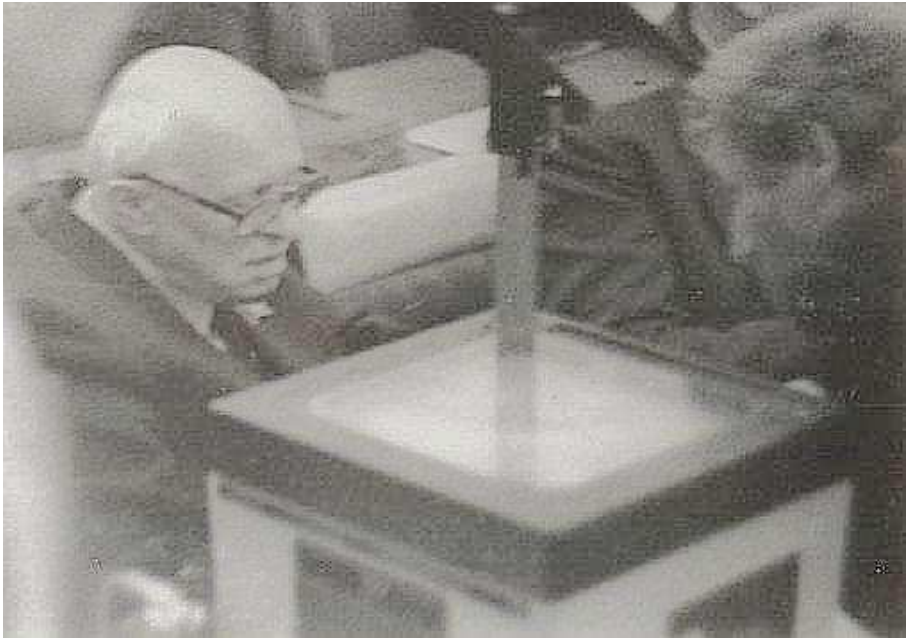}
\end{center}
\caption{One of the last photographs of A.D.Sakharov.}
\label{Sakharov_small.jpg}
\end{figure*}

It is appropriate to start this section with one of the last photographs of
A.D.Sakharov (see Fig.\ref{Sakharov_small.jpg}). It shows the intermission in 
the meeting
chaired by Sakharov at which the present author (among other enthusiastic
speakers) argued that if primordial cosmological perturbations were generated
quantum-mechanically, then the result would be not just something,
but very specific quantum states known as
squeezed vacuum states, and why this should be important observationally. The
notions of the vacuum, a squeezed vacuum and a displaced vacuum (coherent 
states) sounded suspicious to the audience, but Sakharov remained silent.
At some crucial point he astonished me by the question ``which variable
specifically is squeezed ?". Such a question can be asked only by someone who
is perfectly well familiar with the discussed subject and deeply understands
its implications.

\begin{figure*}[t]
\begin{center}
\includegraphics[width = 12cm]{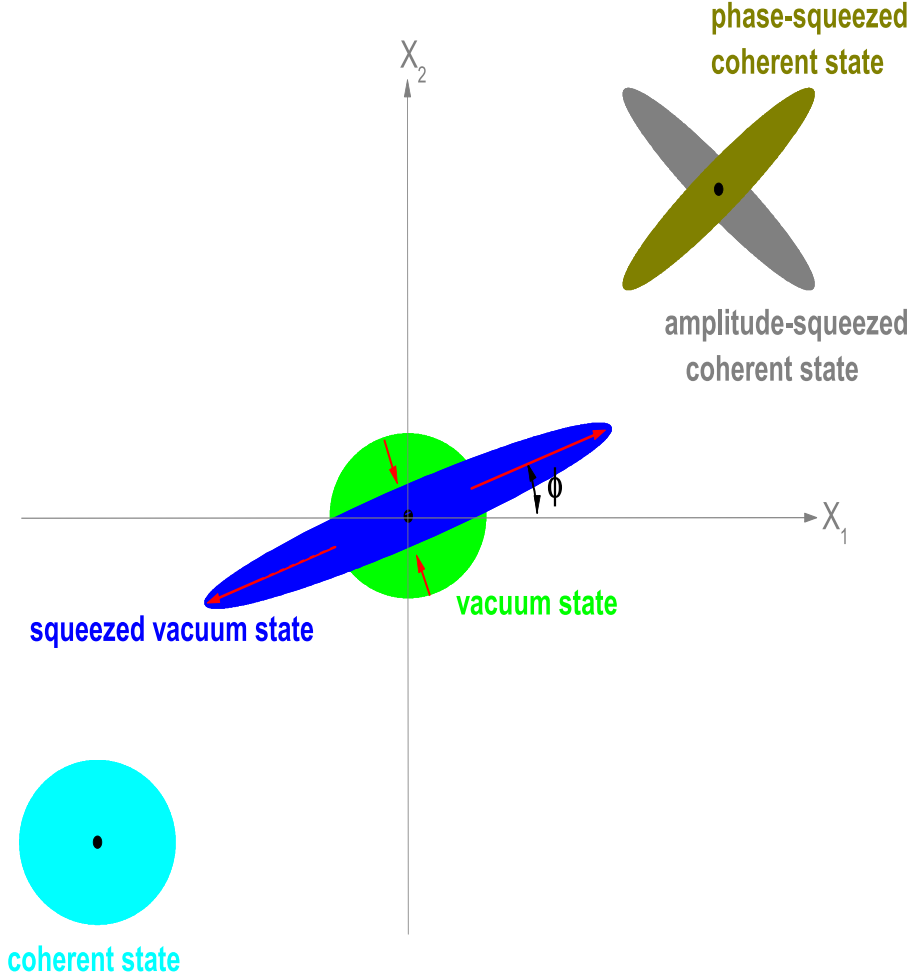}
\end{center}
\caption{Some quantum states of a harmonic oscillator.}\label{fG333.pdf}
\end{figure*}

Indeed, from the sketch in Fig.\ref{fG333.pdf} one can see that simple
quantum states of a harmonic oscillator can greatly differ in mean values
and variances
of conjugate variables. For example, squeezed coherent states can be squeezed,
i.e. have very small uncertainties, either in the number of quanta or in the
phase. This leads to different observational results. I was glad to answer
Sakharov's question, because a squeezed vacuum state can be squeezed only in
phase. The arising correlation of the ${\bf n}$ and ${\bf -n}$ modes is
equivalent to the generation of a standing wave (a two-mode squeezed vacuum
state, more details are given in \cite{GrLH} and \cite{GrWhel}). 
The appearance of the
standing-wave pattern is not surprising if one thinks of the generating
process as the creation of pairs of particles with equal energies and oppositely
directed momenta. Moreover, the phase, almost free of uncertainties in
strongly squeezed vacuum states, smoothly depends on $n$, as the oscillators 
with
different frequencies $n$ start free evolution (rotation of a higly squeezed
ellipse in the $X_1, X_2$ plane) after the completion of the process of
generation (squeezing of the vacuum circle into an ellipse). This provides the
prerequisites for the future Sakharov oscillations.

The generation of excitations in physically different degrees of
freedom -- relic gravitational waves and primordial density
perturbations -- is described by essentially the same equations. The
equation for gravitational-wave mode functions is
\begin{equation}
\label{heq}
h^{\prime \prime} +2\frac{a^{\prime}}{a} h^{\prime}+n^2 h=0,
\end{equation}
while the equation for metric perturbations describing the density
perturbation degree of freedom is
\begin{equation}
\label{zheq}
\zeta^{\prime \prime} +2\frac{(a \sqrt \gamma)^{\prime}}{a \sqrt \gamma}
\zeta ^{\prime}+n^2 \zeta=0,
\end{equation}
where the variable $\zeta(\eta)$ is also known as a curvature perturbation.
Surely, equations (\ref{heq}), (\ref{zheq}) can also be written in the form
of the ``master equation", Eq.(\ref{master}). The function
$\gamma(\eta) \equiv 1 + \left(a/a^{\prime}\right) ^{\prime}$ in
Eq.(\ref{zheq}) is not the constant $\gamma$ that Sakharov \cite{Sakh65} uses
in the equation of state, but for simple equations of state the
scale factor $a(\eta)$ is a power-law function and $\gamma(\eta)$ is then a
constant. In this case,
equations (\ref{heq}) and (\ref{zheq}) are identically the same, and they
have general solutions in terms of the Bessel functions.

The two-mode Hamiltonian
\begin{eqnarray}
\label{hamilt}
   H = nc_{\bf n}^{\dag} c_{\bf n}
     + nc_{-{\bf n}}^{\dag} c_{-{\bf n}}
     + 2\sigma (\eta ) c_{\bf n}^{\dag} c_{-{\bf n}}^{\dag}
     + 2\sigma^\ast (\eta ) c_{\bf n} c_{-{\bf n}} \quad
\end{eqnarray}
is common for these two degrees of freedom, with the coupling function
$\sigma(\eta) = (i/2)[a^{\prime}/a]$ for gravitational waves and
$\sigma(\eta) = (i/2)[(a \sqrt{\gamma})^{\prime}/(a \sqrt{\gamma})]$ for
density perturbations. The coupling functions coincide if
$\gamma(\eta) = const$. As a result of the Schrodinger evolution, the initial
vacuum state of cosmological perturbations (ground state of the
corresponding time-dependent Hamiltonian) evolves into a two-mode squeezed
vacuum (multi-particle) state. In other words, cosmological perturbations are
quantum-mechanically generated as standing waves \cite{GrLH}, \cite{GrWhel}.

The simplest models of the initial stage of expansion of the Universe are
described by power-law scale factors $a(\eta)$. (The four cases of the initial
stage considered by Sakharov \cite{Sakh65} also belong to this category.) Such
gravitational pump fields $a(\eta) \propto |\eta|^{1+ \beta}$ generate
gravitational waves (t) and density perturbations (s) with approximately
power-law primordial spectra:
\begin{eqnarray}
\label{PsPt}
P_{t}(k)=A_t \left(\frac{k}{k_0}\right)^{n_t},~~~
P_{s}(k)=A_s \left(\frac{k}{k_0}\right)^{n_s-1},
\end{eqnarray}
where $n_s-1 = n_t = 2(\beta+2)$, and we are using $k_0=0.002$Mpc$^{-1}$.
The amplitudes $(A_t)^{1/2}$ and $(A_s)^{1/2}$ are independent unknowns, but
according to the theory based on Eqs. (\ref{heq}), (\ref{zheq}) and
(\ref{hamilt}) they should be of the same order of magnitude: $(A_s)^{1/2}  
\sim (A_t)^{1/2} \sim H/H_{Pl}$, where $H$ is the Hubble parameter at the 
initial stage of expansion. [The inflation theory also uses the same 
superadiabatic (parametric) amplification mechanism, which was originally 
worked out for gravitational waves \cite{gr74}, \cite{GrWhel}. However, after 
blind wanderings between variables and gauges, inflationists arrived at what 
they call the ``standard", or even ``classic", result of inflationary theory. 
Namely, the prediction of arbitrarily large $A_s$ in the limit of  
Harrison-Zeldovich-Peebles spectrum $n_s =1$,  and, moreover, for any  
strength of the generating gravitational field, i.e. for any value of the 
Hubble parameter $H$ of inflationary de Sitter expansion $\dot {H} =0$.] 
It is common to 
characterize the contribution of gravitational waves to the CMB by the ratio 
$r \equiv A_t(k_0)/A_s(k_0)$.

Our analysis \cite{ZG} of the 7-year Wilkinson Microwave Anisotropy
Probe data (WMAP7) has resulted in $r=0.285$ and $r=0.2$ as the respective 
maximum likelihood
values in 3-parameter and marginalized 1-parameter searches.
The uncertainties are still large, so these numbers can only be regarded as
indications of a possible real signal. The relic gravitational waves are very
difficult to register but they are the cleanest probe of the very early
Universe \cite{gr74}, \cite{grLett76}, \cite{GrWhel}. 
This is why they are in the centre of several programs aimed at
their identification. The Sakharov oscillations are an element of the whole
picture of quantum-mechanically generated cosmological perturbations, and 
hence the detection of relic gravitational waves would be a huge support 
for the entire theoretical framework.

\section{\label{sec:planck}Expected results of the ongoing observations.
Conclusions.}

\begin{figure*}[t]
\begin{center}
\includegraphics[width = 12cm]{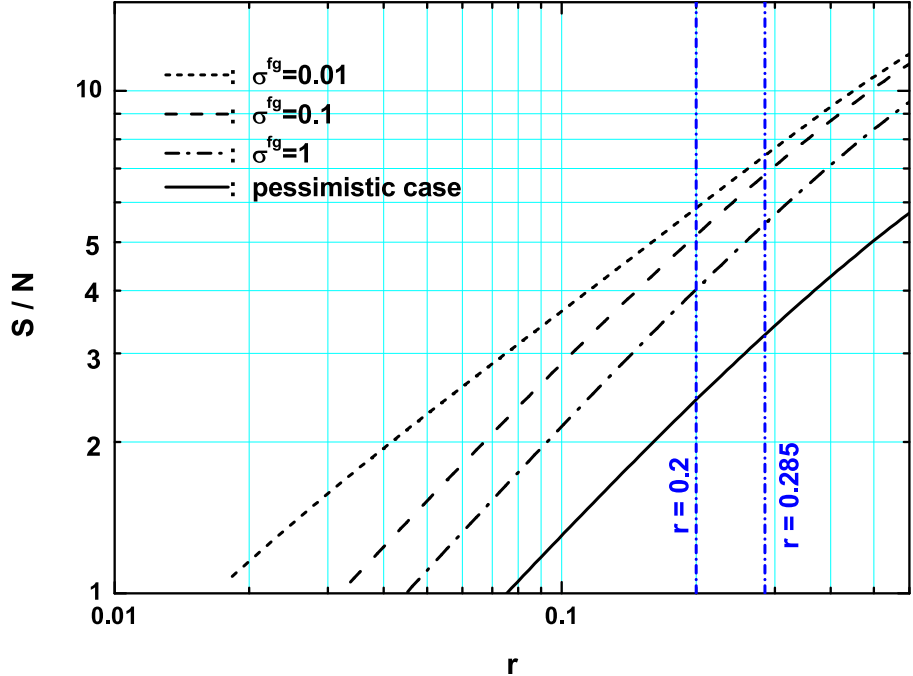}
\end{center}
\caption{The expected $S/N$ ratio in the detection of relic gravitational 
waves by the Planck mission.}\label{nbi_fig4.pdf}
\end{figure*}

The prospects of measuring relic gravitational waves with the help of data
from the currently operating Planck mission appear to be good. In
Fig.\ref{nbi_fig4.pdf}, taken from \cite{ZG}, we show the expected
signal-to-noise
ratio with which the signal will be observed assuming that the indications
found in WMAP7 data are real. A big obstacle is the foreground contamination
which should be carefuly dealt with. The ability, ranging from excellent to
none, of removing contamination is parameterized by the parameter
$\sigma^{\rm fg}=0.01, ~0.1, ~1$. We are also working
with the pessimistic case, in which $\sigma^{\rm fg}=1$ and the nominal
instrumental noise in the $BB$ polarization channel at each frequency is
increased by a factor of 4. We see from the figure that the $S/N$ ratio 
can be as large as $S/N \approx 6$,
and even in the pessimistic scenario it remains at the interesting level 
$S/N>2$.

As was already mentioned above, the planned dedicated observations (e.g. 
\cite{Spider}) may even be able to outline the first cycle in the oscillatory
power spectrum of the gravitational wave background.

In general, we can conclude that the originally proposed Sakharov
oscillations, as well as related phenomena whose existence can be traced back
to the earliest moments of our Universe, are right in the focus of current 
fundamental research.

\section*{Acknowledgements}
The author is gratful to Dr. Wen Zhao for help.

%%%%%%%%%%%%%%%%%%%%%%%%%%%%%%%%%%%%%%%%%%%%%%%%%%%%%%%%%%%%%%%%%%%%%%%%%%
%%%%%%%%%%%%%%%%%%%%%%%%% BIBLIOGRAPHY %%%%%%%%%%%%%%%%%%%%%%%%%%%%%%%%%%%
%%%%%%%%%%%%%%%%%%%%%%%%%%%%%%%%%%%%%%%%%%%%%%%%%%%%%%%%%%%%%%%%%%%%%%%%%%

\end{document}